\documentclass[12pt,aps,prd,preprint,showkeys]{revtex4}
\usepackage[utf8x]{inputenc}
\usepackage{amsmath}
\usepackage{amsfonts}
\usepackage{amssymb}
\usepackage{graphicx}

\begin{document}
\title{Temperature fluctuations and Tsallis statistics in Relativistic Heavy Ion collisions.}
\author{ Abhisek Saha}
 \affiliation{School of Physics, University of Hyderabad, Gachibowli, Hyderabad, India 500046}
\author{Soma Sanyal}
\affiliation{School of Physics, University of Hyderabad, Gachibowli, Hyderabad, India 500046}

\begin{abstract}

We study temperature fluctuations in the initial stages of the relativistic heavy ion collision using a multiphase transport model. We consider the plasma in the initial stages after collision before it has a chance to equilibrate. We have considered Au + Au collision with a center of mass energy of 200 GeV.  We use the non extensive Tsallis statistics  to find the entropic index in the partonic stages of the relativistic heavy ion collisions. 
We find that the temperature and the entropic index have a linear relationship during the partonic stages of the heavy ion collision. This has already been observed in the hadronic phase.  A detailed analysis of the dependence of the entropic index on the system shows that for increasing space time rapidity, the entropic index of the partonic system increases. The entropic index also depends on the beam collision energy. The calculation 
of the entropic index from the experimental data fitting of the transverse momenta deals with the hadronic phase. However, our current study shows that the behaviour of the entropic index in the initial non-equilibrium stage of the collision is very similar to the behaviour of the entropic index in the hadronic stage.

\keywords{temperature fluctuations, Tsallis entropy.}

\end{abstract}

\maketitle

\section{Introduction}

The Relativistic Heavy Ion Collision experiments, both at CERN and at BNL, attempt to understand the thermodynamic properties of strongly interacting systems through various methods \cite{thermo,RR}. Due to the nature of the collision, the fluid immediately after collision constitutes a non-equilibrium system \cite{Kurkela}. The geometry of the collision indicates that a large amount of angular momentum is present in the region of the peripheral collisions which causes rotational motion of the fluid \cite{Jiang}. The motion of the fluid is usually modelled by hydrodynamical simulations or by transport models \cite{Elena}. For any of these simulations, the  critical input is the initial conditions. It is well known that there are fluctuations in the initial conditions which lead to specific structures in the final hadronic spectra. Most of these are number density fluctuations \cite{numberfluc}. Though hotspots and coldspots are known to arise in the rotating fluid \cite{hotspots}, the temperature fluctuations are difficult to study experimentally. 
This has led to most of the efforts being concentrated only on density fluctuations while analysing the fluid dynamics of the system.

 Recently, some groups have been looking at hotspots and temperature fluctuations in these systems \cite{temp}. Temperature fluctuations can help us understand various thermodynamic properties of a system. The time evolution of these temperature fluctuations have been studied as smaller subsystems \cite{bhatta} which are like canonical ensembles with different temperature values. It has been shown that the size of the hotspots do not affect the integrated observables and some of the differential observables. However, there are some other differential observables (such as sub leading principal components) which may be sensitive to these fluctuations \cite{sizehotspots}. In this work, we look at the temperature fluctuations from a different perspective. We use a multiphase transport (AMPT) model to study the temperature fluctuations in the initial stages of the heavy ion collisions. 
Though temperature fluctuations have been studied in the context of transport models, they have mostly been in the Gibb's -Boltzmann statistics. In this paper, we use the non-equilibrium Tsallis statistics to study the temperature fluctuations in the partonic stages of heavy ion collisions.  The extended thermodynamics of Tsallis has been applied previously to non-equilibrium systems \cite{tsallis}. The Tsallis statistics is a generalization of the Boltzmann-Gibbs thermodynamic approach to non-equilibrium systems \cite{Cleymans}. The thermodynamics is characterized by a parameter $q$ which is called the entropic index. For $q = 1$, we obtain the standard Boltzmann-Gibbs statistics. The value of the entropic index is a measure of the departure of the system from equilibrium statistics. It is possible to determine the value of the entropic index from temperature fluctuations.  The Tsallis entropy has been used recently to model various non-equilibrium systems including the relativistic heavy ion collision experiments \cite{tsallisrhic}. The entropic index has also been obtained by fitting the transverse momentum data \cite{compare, Zheng}.

In relativistic heavy ion collisions,  the system takes a finite time to equilibrate \cite{wong}. So, it is reasonable to use the non equilibrium Tsallis statistics to study the system before it reaches a state of equilibrium. In this paper, we have calculated the temperature in the partonic stages  and use the Tsallis statistics to understand the thermodynamics of the collisions in the initial stages. We use the AMPT model to obtain the velocities and positions of the particles at $\tau = 1 fm/c$. We find that during this period the temperature fluctuations obtained are quite significant. We use the Tsallis statistics to find the entropic index from the temperature fluctuations. All the results that we have obtained in this work are within  $\tau= 1 fm/c$ except the time dependent plots in fig 1 and fig 7. Though the exact time of equilibration is not known definitely, it is generally considered to be equilibrated only after $1 fm/c$.

We find the value of the entropic parameter for different rapidities and different centralities and at different collision energies for the partonic stage. The experimental analysis which make use of the Tsallis statistics are in the hadronic phase so the value of the entropic parameter does not match exactly with the values obtained by fitting the transverse momentum data \cite{compare}. They are mostly of the order of $q \sim 1.12$ whereas the values of entropic index that we obtain is mostly greater than $1.2$. We show that the general nature of the change in entropic index with different parameters of the system are in accordance with the observations in the hadronic phase as seen by fitting the Tsallis distribution to the transverse momentum data \cite{compare}. So we conclude that the change in the entropic index with the system parameters is  similar for the partonic and the hadronic phases.  

In section II, we describe the AMPT simulations used to generate the initial partonic stage of the collision. 
In section III, we describe the temperature hotspots obtained in the initial stages of the relativistic heavy ion collisions using the AMPT simulations. Section IV discusses the application of Tsallis entropy to the temperature profiles obtained from the energy density and the calculation of the entropic index from our simulations. We briefly discuss the calculation of the Tsallis entropic index from experimental data and show that it is  close to the entropic index we have found from our simulations in section V. We summarize and conclude the results of this paper in section VI.

\section{The AMPT model and details of our simulation}
The AMPT is a publicly available code for simulating the relativistic heavy ion collision \cite{ampt}. It consists of different components such as the Heavy Ion Jet Interaction Generator(HIJING) model, the Zhang's Parton Cascade (ZPC) model, the Lund string
fragmentation model and the A Relativistic Transport(ART) model for modelling the relativistic transport of the hadrons. The HIJING model is responsible for generating the initial conditions in this simulation. The ZPC models the scattering amongst the partons in the partonic stage. The details of the model may be obtained from ref\cite{ampt}. It has been used quite extensively to study and model the relativistic heavy ion collisions. In this paper we have used the string melting version of the AMPT model. In a previous work, we have already studied in detail, vorticity generation using the AMPT code \cite{abhi}. In this paper, we are using this model again, as it can give the initial state pre-equilibrium fluctuations of the collisions. This choice also arises from the fact that the AMPT has reproduced the  transverse momentum spectra and elliptic flow for low-$p_{T}$ pions and kaons in central Au + Au collisions at collision energies of 200 GeV \cite{lin}.
 
In this paper, we concentrate on the partonic stages. In the AMPT model that we are using, the initial collision results in the formation of strings after partonic interactions. Since we are using the string melting version, 
these strings are converted into soft partons. These are subsequently hadronized by a simple quark coalescence model. The AMPT gives as an output, the particle records of the partons for each event at kinetic freeze out. We can thus obtain the space-time position, the three momentum and the energy of all the partons generated at an event. 

Since we are interested in studying the temperature fluctuations in the system, we generate a grid in the x-y plane, with the cell size such that each cell has a significant number of particles in them. We then calculate the average energy in each cell from the momentum and energy of all the particles in the cell. So in essence, we have divided our whole system into smaller subsystems and we will assume local thermodynamic equilibrium in these smaller subsystems. The system as a whole is considered to be out of equilibrium.   

Since this is a statistical model, we generate a large number of events with the same parameters and initial conditions and then average over all the events to obtain the final values. We have varied the cell size and checked that it does not have any significant contribution to the final values of the temperatures obtained.  We discuss this further in the next section where we discuss the mathematical formulation used to calculate the temperature in these grid cells.

\section{Temperature hotspots in the initial stages}
 
 As mentioned in the previous section, we use the AMPT code to obtain the positions and velocities of the particles at the initial stages of the collision. The energy density distribution function in the AMPT can then be calculated from \cite{zhang}, 
\begin{equation}
\epsilon(x,y) = \sum N_i exp [- \frac{(x-x_i)^2 + (y - y_i)^2 }{2 \sigma^2}]
\end{equation}  
We set the  Gaussian width  at $\sigma = 0.5$ fm. We also have $N_{i} =\frac{N}{2 \pi} (\frac{1}{\sigma^{2}\tau}) E_{i}$. $N$ is the normalization vector. $(x_i, y_i, z_i)$  are the position coordinates and $E_{i}=\sqrt{\textbf{\textit{p}}_{i}^{2} + m_{i}^{2}}$ is the energy value of the $i$-th parton and the sum in eq.(1) runs over all the particles in an event. We have taken a rapidity window $-3<\eta<3$ where $\eta=\frac{1}{2}ln\frac{t+z}{t-z}$ is the space-time rapidity and $\tau=\sqrt{t^{2}-z^{2}}$. We have calculated the energy density taking into account all the four components of the energy momentum tensor.  As mentioned in the previous section, we have divided the whole system into smaller subsystems with grid sizes $dx = dy = 0.3 fm$. We have also changed the size of the grid cells and checked with values of grid cell sizes ranging from $0.1 fm$ to $0.5 fm$, we have found that our results do not depend upon the cell size as long as it is within this range. In each of these grid cells, we have assumed local thermal equilibrium and used the ideal gas, Gibbs- Boltzmann statistics to calculate the temperature using the equation, 
\begin{equation}
\epsilon (x,y) = 12 ( 4 + 3 N_f) (\frac{T^4}{\pi^2})
\end{equation}
Here $N_f$ is the number of quark flavours; we have taken $N_f=3$.  

This equation for calculating the temperature is based on equilibrium statistical mechanics. The definition of temperature in non-equilibrium systems is still not well defined. This is because, there are problems in extrapolating the zeroth and the second laws of thermodynamics to non-equilibrium systems. In general, in equilibrium thermodynamics, an equilibrium system can be divided into subsystems and the temperature registered by all the subsystems will be exactly the same. However, for an out of equilibrium scenario, all the subsystems may not register the same temperature. Therefore, it is difficult to define one single temperature in an out of equilibrium scenario \cite{casas}. The Tsallis statistics which we will be using later to understand the departure from equilibrium for this system has two parameters, a temperature and an entropic index. The temperature parameter can be related to the Gibbs-Boltzmann temperature if the entropic index of the system is known. Since the entropic index of the relativistic heavy ion collision is not known a priori, the temperature parameter in the Tsallis distribution cannot be calculated. However, for entropic index values close to one, the Tsallis parametric temperature will be approximately equal to the equilibrium Gibbs-Boltzmann temperature since the Tsallis statistics goes to the Gibbs- Boltzmann statistics when the entropic index is equal to one. So we can conclude that the temperature used in the Tsallis entropy will be approximately equal to the equilibrium temperature of the system. We calculate the equilibrium temperature in each of the small cells of the grid that we have mentioned before. In the previous section, we have mentioned that we have divided the system into small grids, each of these grid cells is like a subsystem. Thus, the equilibrium temperature that we obtain for each cell is not the temperature of the system as a whole but the temperatures of the small subsystems that we have divided the space into. Since the system as a whole is a non equilibrium system, the temperature of these subsystems will not be the same and will reflect the temperature fluctuations of the system. The temperature so obtained is plotted in the $x-y$ plane. We have done both event by event as well as event averaged plots. We first discuss the event by event plots which show the initial fluctuations at different times and collision energies.   
\begin{figure}
\includegraphics[width = 13cm, height=9cm] {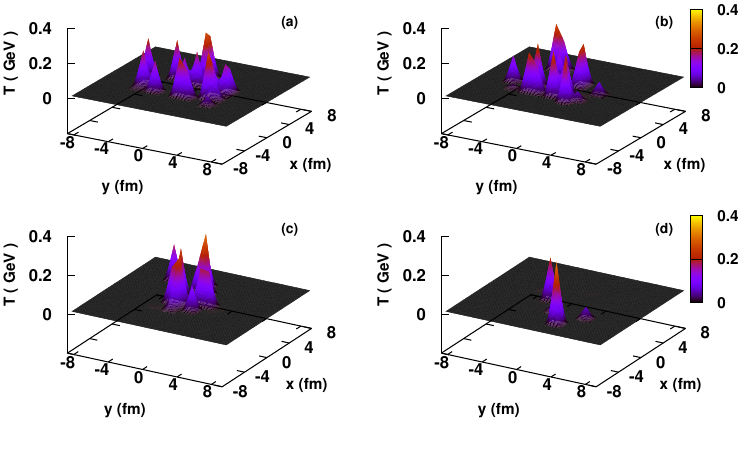}
\caption{Temperature fluctuations at times a) 1 fm/c b) 2 fm/c c) 3 fm/c d) 5fm/c at  $\sqrt{s_{NN}} = 200$ GeV and $-1<\eta<1$ for Au + Au collisions }
\end{figure}
Fig. 1 shows the temperature fluctuation at different times. We see that the temperature fluctuations are higher in the initial stages but decrease with time.  
\begin{figure}
\includegraphics[width = 13cm, height=9cm] {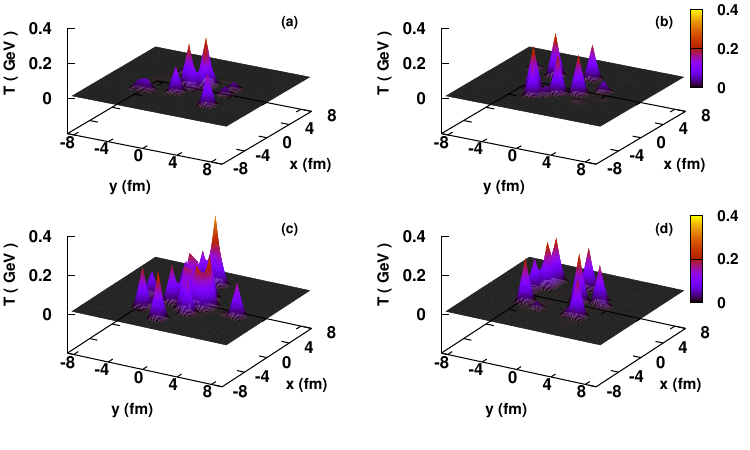}
\caption{ Temperature fluctuations at different $\sqrt{s}$ values a)  $\sqrt{s_{NN}} = 19.6$ GeV   
b) $\sqrt{s} = 62.4 $ GeV    c)  $\sqrt{s} = 100$ GeV   d) $\sqrt{s} = 200$ GeV at time $\tau = 1$ fm/and $-1<\eta<1$ for Au + Au collisions }
\end{figure}
Fig. 2 shows the temperature fluctuations for different collision energies. As is seen, the general pattern of the fluctuations remain the same irrespective of the collision energy. Only the amplitude of the fluctuations increases at higher collision energies. Such temperature fluctuations have been observed before in previous studies \cite{hotspots,basu}. In the first study, the temperature fluctuations are studied  in the same way as temperature fluctuations are  studied for the CMBR (Cosmic Microwave Background Radiation) in the early universe. The second study concentrated on finding the specific heat of the system from temperature fluctuations. We, on the other hand, treat the fluid as a non-equilibrium system and though the temperature hotspots appear similar at different times, we would like to analyse them further to find the differences at various collision energies. So in our case, the equilibrium description of temperature is used to find the temperature of the smaller subsystems within the larger system. This means that we have assumed local thermal equilibrium in the small regions that make up the complete system. Since these different regions do not have the same temperature, it means that the system as a whole is out of equilibrium \cite{casas}.

\section{Tsallis entropy and the entropic index}
 
As mentioned before, the Tsallis statistics is  characterized by the non-extensivity parameter $q$, with $|q-1|$ being a direct measure of the temperature fluctuations \cite{wilk}.
Though initially, it was used to fit the hadron data at different collision energies \cite{cleymans1}, it was subsequently used in many contexts of the relativistic heavy ion collision. In some papers, it has been argued that the Tsallis statistics can only be used for the hadronic data fitting and is not appropriate for thermodynamical or theoretical understanding \cite{simon} of the statistical mechanics of the system. However, there are various frameworks in QCD where the Tsallis statistics is still being discussed and applied. A detailed discussion on the various application of Tsallis statistics to the heavy ion collision is available in the introduction to ref.\cite{barboza}. As mentioned in ref.\cite{barboza}, apart from the hadrons, the Tsallis statistics has also been applied to quark matter \cite{deppman}. In all these cases the thermodynamics of the system has also been studied. Generally, to derive these thermodynamic properties, the partition function of an ideal gas is used. For the phase diagram, the analysis shows that the critical temperature in the phase diagram using Tsallis statistics is generally lower than that obtained in the corresponding Boltzmann-Gibbs statistics\cite{barboza}. This may happen as the temperature obtained using the equilibrium definition will always be higher than the temperature obtained using any other method \cite{casas}.

Extensions of the MIT bag model using the Tsallis statistics have also been discussed in ref.\cite{deppman}. In this case, the Equation of State (EoS) has been studied for different values of the bag parameter and the entropic index. Here though the shape of the phase diagram remains the same for both the statistics, the critical temperature is found to decrease with  increasing values of the entropic index.  The Tsallis entropy formula has also been obtained in ref.\cite{biro} from a thermodynamic system comprising of a reservoir and a subsystem. The temperature is defined and the entropic index is found to be related to the heat capacity of the reservoir. We mention these particular references as we are interested in the relationship between the temperature and the entropic index in the initial stages of the heavy ion collision. This is the quark gluon plasma phase, in almost all other cases, the Tsallis statistics has been applied to the hadronic phase only. In ref \cite{sukanya}, however both the hadronic and the quark gluon plasma case have been discussed theoretically.  

In the initial stages of the heavy ion collisions, as we have mentioned before, temperature fluctuations have been observed. It is these temperature fluctuations which are of interest to us. 
The relation between the entropic index $q$ and Tsallis entropy in a system with fluctuating temperature has been discussed previously in the literature \cite{beck}. The quantity that is used here is $\beta$, the inverse of temperature. If a non-equilibrium system is formally described by a fluctuating $\beta$, then the generalized distribution functions of non-extensive Tsallis statistics are a consequence of integrating over all possible fluctuating $\beta$'s provided that the $\beta$ is $\chi^2$ distributed \cite{beck,parvan}. 

Since our system too has temperature fluctuations, we fit the probability distribution of $\beta$ (i.e the temperature inverse)  with a       $\chi^2$ distribution similar to ref.\cite{beck}. In ref.\cite{beck}, the author has shown that for any system with fluctuating temperatures,
the following distribution can be used to obtain the relation between the entropic index and temperature, 
\begin{equation}
f(\beta) = \frac{1}{\Gamma(\frac{1}{q-1})} \left(\frac{1}{(q-1)\beta_0} \right)^{(\frac{1}{q-1})} \beta^{\frac{1}{q-1} - 1}  exp \left(\frac{-\beta}{(q-1)\beta_0}\right)
\end{equation}
As we can see the function itself has the dimension of temperature $(\beta)^{(-1)}$, while the entropic index $q$ is dimensionless.
\begin{figure}
\includegraphics[width = 8.6cm,height=6.4cm]{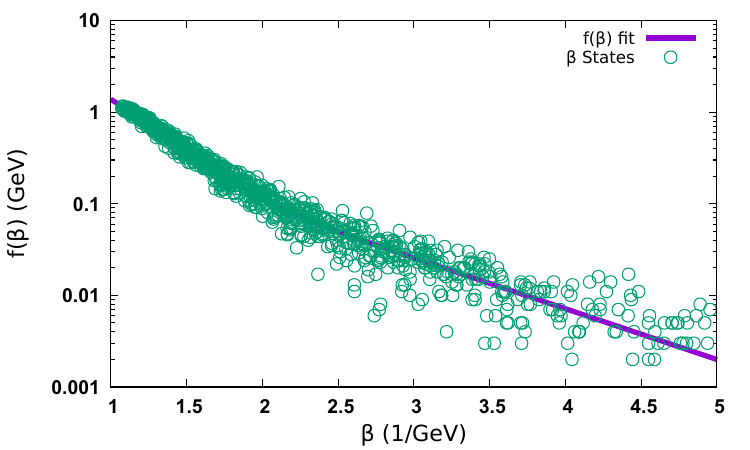}
\caption{The plot of $f(\beta)$ for the temperature fluctuations at a collision energy of $\sqrt{s} = 200 $ GeV (Au - Au collision, $|\eta|<3$). The green circles are the fluctuating states. The value of $q$ is obtained by fitting a $\chi^2$ distribution to the plot. Purple line shows the fitted curve. The final values of the two parameters $q$ and $\beta_0$ are $q = 1.51158$ and $\beta_0 =  0.968718 GeV^{-1} $ , with the asymptomatic standard error for $q$ being $3.5\%$ and for $\beta_0$ being $0.33\%$. }
\end{figure}
We obtain $\beta$ from our simulations and after fitting with the $\chi^2$ distribution, the plot is shown in fig 3.  We obtain a good fit for our temperature fluctuations and the entropic index can be obtained from these fits. The constant $\beta_0$ is the average of the fluctuating $\beta$ \cite{beck}. The value of $q$ is found to be $q = 1.51158 $ and $\beta_0 =  0.968718 GeV^{-1}$ with the asymptomatic standard error for $q$ being $3.5\%$ and for $\beta_0$ being $0.33\%$.  
It has been seen in most physical systems that the entropic index from the Tsallis statistics has some dependency on system parameters. For example, the value of $q$ is observed to depend on the spatial scale.  In the next section, we study how the entropic index or $q$ value changes with various parameters of the system.

\section{Results and Discussions}

Our first result is the relationship between the temperature and the entropic index. We have obtained the entropic index by fitting the $\chi^2$ distribution to the inverse of temperature ($\beta$). We find that there exists a linear relationship between the temperature and the entropic index which can be fitted with a straight line. This is shown in fig 4. 
\begin{figure}
\includegraphics[width = 8.6cm,height=6.4cm]{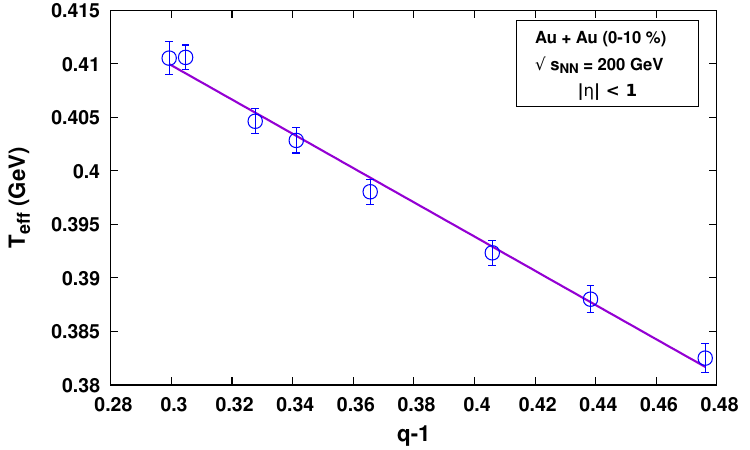}
\caption{The plot shows the dependence of the effective temperature $T_{eff}$ in GeV on the values of $q$ (dimensionless) at a collision energy of $\sqrt{s_{NN}} = 200 $ GeV for the Au+Au system, $(0-10\%)$ centrality with $|\eta| < 1$. } 
\end{figure} 

Recently, there have been attempts to obtain the $(q-1)$ values from experimental data \cite{compare}.In that approach the $T_{eff}$ is defined differently. The dependence of the effective temperature $T_{eff}$ on the parameter $q$ has been studied for negative pions and antiprotons in different reactions\cite{wilk}. The data can be fitted with a straight line whose slope depends on the reactions and the particles chosen. This has been shown in ref.\cite{de}, where the authors have analysed data from the P + P, Au + Au and D + Au collisions, all of them at 
 $\sqrt{s_{NN}} = 200 GeV$ at RHIC-BNL. The entropic index values that they have got for the positive pions are in the range of $1.12$, whereas the values for entropic index that we have got is greater then $1.28$. The main difference with the previous work and our work is that we are dealing with the initial stages of the heavy ion collisions, hence we do not have any hadrons.  As mentioned before we have obtained the temperature from the AMPT simulations and the corresponding values of the entropic index $q$ for the partonic stages before equilibration. We find that even in our case, a linear  dependence on the $q$ values can be obtained. The data from the simulation can be fitted with a straight line. The slope we get is however smaller than the slope obtained for the hadronic particles. 

The experimental data fitting in ref \cite{biro1} has already indicated that the $T_{eff}$ vs $q$ dependency may not be the same for all systems. While for $Au$ +$Au$ collisions, the slope we obtained is  similar to ref. \cite{wilk}, it was shown in \cite{biro1} that the $p-p$ collision at high collision energies has a different slope. For the $Pb-Pb$ collision in the same reference they have obtained a different slope. Thus though we get a linear dependency between $T_{eff}$ and $(q-1)$, the slope of the straight line depends on the particular system being studied as well as the collision energy. 

The definition of the temperature is also very crucial to the system. As mentioned in ref.\cite{wilk}, the effective temperature is the temperature which occurs when there is both the fluctuations of the temperature $T$  and some energy transfer
taking place between the source and the surroundings. Thus the effective temperature is not the thermodynamic temperature of the system. In our case, we have divided the system into smaller subsystems and have assumed that there is local thermal equilibrium in these smaller spaces. The system has a whole does not have the same temperature through out as it is out of equilibrium.  Since the system at this stage is more out of equilibrium than the system in the hadronic phase, hence the values of entropic index is higher in our case as compared to the values of the entropic index obtained from previous studies \cite{de}.  In ref.\cite{barboza}, the Tsallis entropy has been applied to the MIT bag model and the relationship between the temperature and the energy density comes out to be 
\begin{equation}
\epsilon = [\frac{7}{4} g_Q + g_G]\frac{\pi^2}{30} T^4 +  \frac{8 \pi^2}{30}g_Q g_G \frac{\pi^2}{90}(q-1) T^7. 
\end{equation}
Here $g_Q$ and $g_G$ are the quark and gluon degrees of freedom. So in addition to the $T^4$, term there is a second term which has the entropic index and is proportional to $T^7$. The temperature obtained from this equation would be different from that obtained from the equilibrium relation $\epsilon = [\frac{7}{4} g_Q + g_G]\frac{\pi^2}{30} T^4$. Thus the entropic index corresponding to the different temperatures will vary. So the exact value of the entropic index calculated will depend upon the definition of the temperature that is used for calculating the entropic index. The entropic index can also be calculated from the temperature fluctuations directly by using the formula, 
\begin{equation}
q = 1 + \frac{Var(T)}{<T>^2}
\end{equation}
Here $Var(T)$ is the variance of the temperature $T$ which is the expected value of the squared deviation from the mean of the temperature. As we can see, the entropic index is dimensionless.  This definition has been used by Wilk et.al. \cite{wilk} and they have obtained similar relationship between the temperature and the entropic index as in fig 4. For different collision energies, we use this definition of the entropic index to make comparisons to previously obtained values of the entropic index \cite{biro1,de}. This is because of its direct relation to the temperature fluctuations. 

Next, we have done a detailed analysis of how the $q$ value changes for different parameters of our system. In fig 5, we show the variation of $q$ for different space time rapidity ($\eta$) values at different collision energies ($\sqrt{s}$). Here we find that with increasing values of space time rapidity, our $q$ value increases plateauing out for higher collision energies. Increasing the space time rapidity means that the system size is also increasing. If we take a larger system size, the departure from equilibrium will be larger too, this will cause an increase in the value of the entropic index.


 We would like to emphasize the point that the value of the entropic index of the system depends crucially on the colliding system. The difficulty in using the Tsallis distribution seems to be that the results are dependent on the produced particles and the colliding particles. A recent paper \cite{biro2} has done a detailed study for a large range of collision energies, a large variety of particles and for various different systems. They show that the $q$ values appear to be proportional to $\sqrt{s}/m$ \cite{biro2}, they have also mentioned that the $q$ value depends on the multiplicity of the particles.
  
The major difference between our work and the others is that we are calculating the $q$ value for the partons whereas in all the previous cases, it is the hadrons which are used to obtain the $q$ values. We find that our values are different from the values obtained from the hadronic spectrum. The values obtained by fitting the transverse momenta of the final particles is of the range of $1.12 $ \cite{compare}, whereas our values start from $1.1$ but increase upto $1.18$ (fig 5).  There are two major reasons why this could have happened. The first we have already discussed previously. It is the definition of the temperature that we have used. If a different Equation of State (EoS) is used, the temperature values and subsequently the entropic index values would also be different. The other reason is the fact that we are looking at the partonic system. It is quite possible that, at the early stages before the phase transition, the system is more likely to be in a strongly non-equilibrium state. The system moves towards equilibrium as it evolves.  After the hadronization, the fluctuations are less and the system moves towards equilibrium once again. So the $q$ values, which are a measure of the departure from the equilibrium of a system, will be different in the partonic state as compared to the hadronic state of the system.

\begin{figure}
\includegraphics[width = 8.6cm,height=6.2cm]{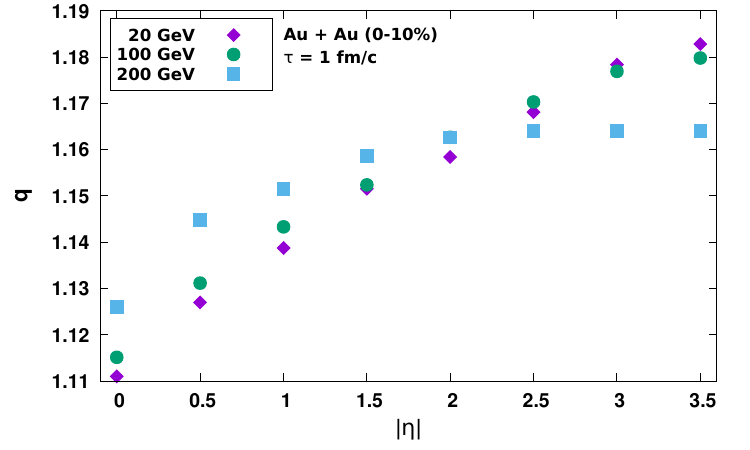}
\caption{The plot shows the variation of $q$ for different space time rapidity ($\eta$) values at different collision energies ($\sqrt{s}$)for Au + Au collision in the (0-10\%) centrality range at $\tau = 1fm/c$. }
\end{figure}

Our next results show that the $q$ values  have a dependency on the beam collision energy. Though the entropic index is not determined directly from the experiments, the values used for fitting the data indicate that the entropic index depends upon the beam collision energy\cite{abelevAlice}.    
In fig 6, we show the variation of $q$ for different $\sqrt{s_{NN}}$ values at different centralities. From both figs 5 and 6, we find that at lower collision energies the Tsallis entropic index is lower. However, the dependence is dependent on the centrality of the collisions. This is shown in fig 6, where we see  the $q$ value, for peripheral collisions (60\% - 80\%) are lower than the $q$ value at central collision (0-10\%) below $100$ GeV. Above $100$ GeV, the opposite is true, i.e  the $q$ value, for  centralities (60\% - 80\%) is higher than the $q$ value at centralities (0-10\%). This result has also been obtained from experimental data in a recent paper \cite{bedanga}. In ref.\cite{bedanga}, the authors have fitted the experimental data from Au + Au collision at the RHIC from the beam energy scan program \cite{STAR} as well as data from the PHENIX collaboration \cite{phenix}. They have also observed the dependency of the $q$ parameter on the centrality and the collision energy.  
The total change in the $q$ value for central collisions is also much smaller than the total change in the peripheral collisions. This seems to indicate that it is not the multiplicity alone that determines the entropic index.

\begin{figure}
\includegraphics[width = 8.6cm,height=6.2cm]{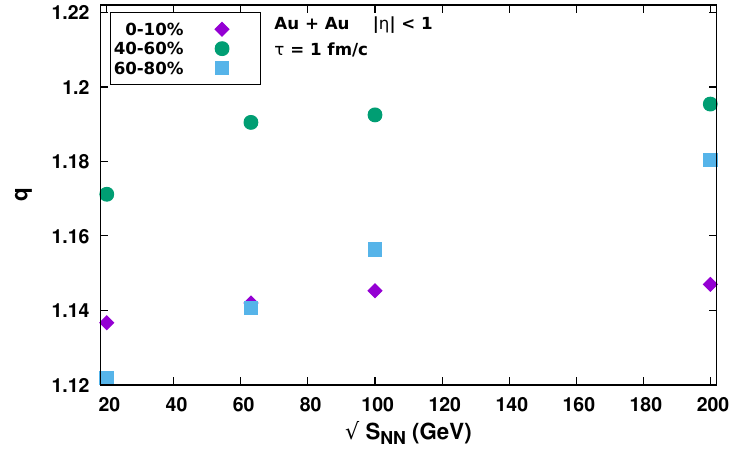}
\caption{This plot shows the variation of $q$ for different $\sqrt{s_{NN}}$ values at different centralities for Au + Au collisions. Here $|\eta| < 1$ and $\tau = 1 fm/c$.}
\end{figure}

 In fig 7, we show the variation of $q$ with proper time ($\tau$)  at different collision energies ($\sqrt{s_{NN}}$). This is not however an evolution of the system with time, it is merely the calculation of the entropic value at different times. 
\begin{figure}
\includegraphics[width = 8.6cm,height=6.2cm]{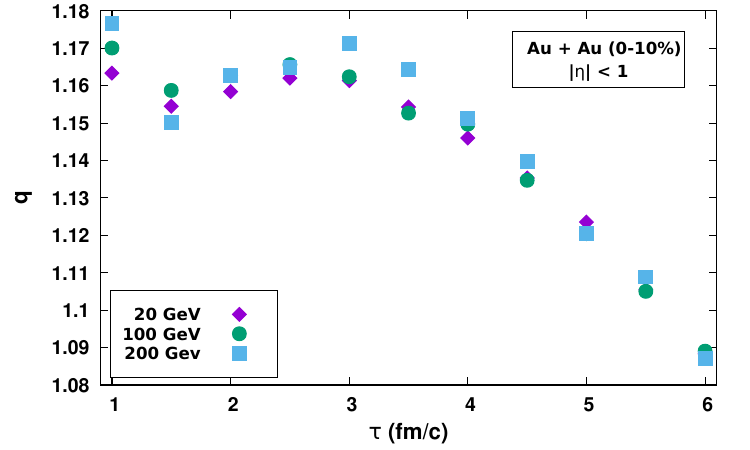}
\caption{This plot shows the variation of $(q)$ with proper time ($\tau$)  at different collision energies ($\sqrt{s_{NN}}$) for Au + Au collisions at $|\eta| < 1$ and for  (0-10\%) centrality.  }
\end{figure}
As can be seen from the graph, the $q$ value appears to peak around $3 fm/c$ and then decreases with increasing $\tau $.  The nature of the graph does not change for different values of collision energy. A lower $q$ value means that the system is close to equilibration. However, since we are using the AMPT model, the increase and subsequent decrease of the $q$ value may be attributed to the increase and decrease in the energy density of the particles. Generally, the energy density of the particles produced in the AMPT model, increases upto $3 -4 fm/c$ and then gradually decreases \cite{xu}. Since the entropic index is calculated from the energy density, hence it shows a similar behaviour.

When the experimental data is fitted with a Tsallis distribution, it has been observed that the $q$ value depends on the nature of the hadrons produced \cite{biro2} as well as the kind of system. Here we do not have any hadrons as we are looking at the initial stages of the heavy ion collision, hence we cannot match the result exactly with the values obtained from fitting the transverse momenta spectra. We find that our $q$ values are close to the values obtained by fitting the transverse momenta spectra in ref.\cite{bedanga}. They also show the same behaviour as the $q$ values obtained by fitting the transverse momenta in different experimental data.  Similar to the analysis in ref.\cite{bedanga}, our results also show a dependency on the collision energy of the particles, however, we have obtained the results from temperature fluctuations. The parameter $q$ is seen to be increasing with collision energy. We also find that there is a dependency on the centrality of collisions.

\section{ Summary and Conclusions}

In conclusion, we have used the AMPT model to simulate the initial stages of the relativistic heavy ion collision. Since in the early stages of the collision the system has a lot of temperature fluctuations, we have calculated these temperature fluctuations in the partonic phase. 
The system is divided into smaller grid cells and the temperature in these smaller subsystems is obtained by using the ideal gas energy density - temperature relationship. For an ideal gas in equilibrium, all the temperatures in this smaller subsystems should have been the same. However, we find that the temperature in the smaller grid cells are not the same throughout the whole system. This indicates that the system is not in equilibrium. We plot the temperature fluctuations of the system in the initial stages and calculate the temperature fluctuations between different cells in the grid. The temperature fluctuations of an out of equilibrium system can be studied using the Tsallis statistics provided the inverse of the temperature $\beta$ can be fitted with a $\chi^2$ distribution. The $\beta$ obtained from our simulations is fitted with a $\chi^2$ distribution and then used to obtain  the entropic index $(q)$.

Our first result was obtaining a relationship between the effective temperature of the system and the entropic index in the partonic stages of the collision. Similar to previous studies, we find that there is a linear dependency between the temperature $T_{eff}$ and the entropic index $q$ for the partonic system. The slope of the line, however, depends on the kind of particles chosen to obtain the temperature of the system. Though there have been previous work showing this relationship, that was in the hadronic phase. We show that a similar relationship also holds in the partonic phase. Since, we have looked at the partonic system in the initial stages of the heavy ion collision, the slope we obtain is different from the hadronic system.   

We have checked the dependency of the entropic index on the space time rapidity, the collision energy and the centrality of collisions. We find that the entropic index increases with an increasing  space-time rapidity. This may be related to the multiplicity of the particles, as higher multiplicity results in a lower $q$ value. However, multiplicity alone cannot explain the variation of the entropic index as we have found by studying the other parameters, the dependence of the entropic index on collision energy and centrality. Our second important result is that in the range of collision energies $20$ GeV - $200$ GeV, we see that the entropic index increases with increasing collision energy in the partonic system.
For the centrality values, the relationship is convoluted with the collision energy. For central collisions, the range of $q$ is nearly constant between $20$ - $200$ GeV.  However, for peripheral collisions, the range of $q$ increases in the same collision energy range. Moreover, in the range of $40-80$ GeV, the $q$ value for central and peripheral collisions are nearly the same. Below this range the $q$ value for central collision is higher than the $q$ value for peripheral collisions. Above this range of $40-80$ GeV, the $q$ value for central collision is lower than the $q$ value for peripheral collisions. These conclusions agree with recent analysis of transverse momentum data using Tsallis statistics for the hadronic stage.  
Our $q$ values are different compared to those that are obtained by experimentally fitting the transverse momentum data \cite{compare}. We attribute this to the fact that at the initial stages of the heavy ion collision, the system has far more temperature fluctuations than in the hadronic stage. The previous analysis in the literature reflects the $q$ values from the hadron spectra, while we have obtained it for the partonic stage, before equilibration. 

Lastly, by analysing the temperature fluctuation at different times, we have found that the 
entropic index changes at different times. The entropic index increases, reaches a peak value and then decreases as time increases. This seems to follow the rise and fall of the energy density as seen in relativistic heavy ion collision. A previous study of the heavy ion collisions using the AMPT and the hadronic resonance gas model has shown that the energy density increases with time, reaches a peak around $3-4$ fm/c and then goes down gradually. This is what is reflected in the change of the entropic index with time. So the change in the entropic index with time is a result of the change of the energy density with time.

  Though there are many different interpretations of the application of Tsallis statistics to the collider experiments, we feel that there are several points that are still not clear. The experimental data fitting also points to the fact that the value of the entropic index depends very crucially on the particles chosen as well as the collision energy of the experiment. The results for $p-p$ collisions and $Au - Au $ collisions are not the same. 
Due to all these reasons, we feel that the role of the entropic index is not well understood in heavy ion collisions.  However, our current work indicates that a non-extensive formalism 
can be used in conjunction with a transport model to study the partonic stages of relativistic heavy ion collisions.  

\begin{center}
 Acknowledgments
\end{center}  
 For computational infrastructure, we acknowledge the Center for Modelling, Simulation and Design (CMSD) at the University of Hyderabad, where part of the simulations was carried out. A.S is supported by INSPIRE Fellowship of the Department of Science and Technology (DST) Govt. of India, through Grant no: IF170627.


\begin{thebibliography}{100}

\bibitem {thermo} V. Chandra, V. Ravishankar, Eur. Phys. J. C 59, 705-714 (2009).

\bibitem {RR} A. Bhattacharyya, P. Deb, S. K. Ghosh, R. Ray, and S. Sur, Phys. Rev. D 87, 054009 (2013).

\bibitem {Kurkela} Kurkela A, Mazeliauskas A, Paquet JF, Schlichting S, Teaney D. , Phys. Rev. Lett. 122, 122302 (2019).

\bibitem{Jiang} Y. Jiang, Z.W. Lin, J. Liao, Phys. Rev. C 94, 044910 (2016).
 
\bibitem {Elena} Yingru Xu, P. Moreau, T. Song, M. Nahrgang, S. A. Bass, and E. Bratkovskaya, Phys. Rev. C 96, 024902 (2017).

\bibitem {numberfluc} K. Paech and A. Dumitru,  Phys. Lett. B 623, 200-207 (2005).

\bibitem {hotspots} \'{A}gnes M\'{o}csy and Paul Sorensen, Nucl. Phys. A 855(1) 241-244 (2011).

\bibitem {temp} Sumit Basu et. al. Phys. Rev C 94, 044 901 (2016);  Sumit Basu et al. AIP Conference Proceedings 1701, 060004 (2016).

\bibitem{bhatta} Bhattacharyya, T., Garg, P., Sahoo, R. et al.  Eur. Phys. J. A 52, 283 (2016).

\bibitem{sizehotspots} Fernando G. Gardim, Frédérique Grassi,  Pedro Ishida,  Matthew
Luzum,  Pablo S. Magalhães,  and Jacquelyn Noronha-Hostler, Phys. Rev. C 97, 064919 (2018).

\bibitem{tsallis} Peter W. Egolf and Kolumban Hutter, Entropy 20(2), 109 (2018).
\bibitem{Cleymans} Azmi, M.D., Cleymans, J. Eur. Phys. J. C 75, 430 (2015).
\bibitem{tsallisrhic} Zebo Tang, Yichun Xu, Lijuan Ruan, Gene van Buren, Fuqiang Wang, and Zhangbu Xu, Phys. Rev. C 79, 051901 (2009)(R).

\bibitem{compare} Rui-Fang Si, Hui-Ling Li, Fu-Hu Liu, Advances in High Energy Physics, vol. 2018, Article ID 7895967, 12 pages, (2018).
\bibitem{Zheng} H. Zheng and Lilin Zhu, Advances in High Energy Physics, Article ID 180491 (2015).
\bibitem{wong}S. M. H. Wong, Phys. Rev. C 54, 2588-2599,(1996).
\bibitem{ampt} Zi-Wei Lin, Che Ming Ko, Bao-An Li, Bin Zhang, and Subrata Pal, Phys. Rev. C 72, 064901 (2005).




\bibitem{abhi} A. Saha, S. Sanyal, Int. J. Mod. Phys. E, Vol 29, No 01.2050001 (2020).

\bibitem{lin}Z. W. Lin, Phys. Rev. C 90, 014904 (2014).
\bibitem{zhang} Y. Zhang, J. Zhang, T. Chen, D. Liu, and Y. Chao, Phys. Rev. C 96, 044914 (2017); Rajeev S. Bhalerao, A. Jaiswal, and S. Pal, Phys. Rev. C 92, 014903 (2015).
\bibitem{casas} J. Casas-V´azquez and D. Jou, Rep. Prog. Phys. 66, 1937–2023 (2003).
\bibitem{basu} Sumit Basu et al. J. Phys.: Conf. Ser. 668, 012043 (2016).

\bibitem{wilk} G. Wilk and Z. Wlodarczyk, Phys. Rev. C. 79, 054903 (2009).
\bibitem{cleymans1} J. Cleymans, G. Hamar, P.Levai and S. Wheaton, Journal of Physics G: Nuclear and Particle Physics, 36 (2009).

\bibitem{simon}A. Simon and G. Wolschin, Phys. Rev. C 97, 044913 (2018).

\bibitem{barboza} Carolina Barboza Mendoza and G. Herrera Corrala, Eur. Phys. J. A 55: 146 (2019).

\bibitem{deppman}E. Megías, D. P. Menezes, and A. Deppman, Physica A: Statistical Mechanics and its Applications 421,  15-24 (2015); P.H. Cardoso, T.N. da Silva, A. Deppman, and D.P.  Menezes, The European Physical Journal A, 53(10), 191 (2017).
\bibitem{biro}T. S. Bir\'{o}, G. G. Barnaf$\ddot{o}$ldi1, P. V\'{a}n, Eur.Phys. J. A, 49:110 (2013) 
\bibitem{sukanya} S. Mitra, Eur. Phys. J. C, 78 1, 66 (2018).
\bibitem{beck} C. Beck, Europhys. Lett., 57 (3), pp. 329-333 (2002).
\bibitem{parvan} A.S. Parvan, Eur.Phys. J. A 53, 53 (2017); A.S. Parvan and T. Bhattacharyya, Eur. Phys. J. A 56, 72 (2020)


\bibitem{de} B. De, S. Bhattacharyya, G. Sau and S. K. Biswas, Int. J. Mod. Phys. E 16, 1687 (2007).
 
\bibitem{biro1} K. Shen, G. G. Barnaf$\ddot{o}$di and T.S. Bir\'{o}, Universe 2019, 5(5), 122; (2019).




\bibitem{biro2} G. B\'{i}r\'{o}, G. G. Barnaf$\ddot{o}$ldi1, T. S. Bir\'{o} and K. Urmossy, arXiv:2003.03278 [hep-ph].
\bibitem{abelevAlice} B. Abelev et al., (ALICE Collaboration), Phys. Lett. B, 717, 162 (2012). 

\bibitem{bedanga}Rajendra Nath Patra, Bedangadas Mohanty, Tapan K. Nayak, arXiv 2008.02559
\bibitem{STAR} L. Adamczyk et al., (STAR Collaboration), Phys. Rev. C 96, 044904 (2017); J. Adam et al., (STAR Collaboration), Phys. Rev. C 101, 024905 (2020).
\bibitem{phenix}S. S. Adler et al., (PHENIX Collaboration), Phys. Rev. C 69, 034909 (2004).

\bibitem{xu}J. Xu, C.M. Ko, Phys. Lett. B 772,  290–293, (2017).





\end{thebibliography}
\end{document}